# Geometric Memory Generates Irreversible Transport in Time-Periodic Irrotational Flows

Mounir Kassmi

University of Tunis El Manar- Rommana, 1068 Tunis, Tunisia

Email: mounirkassmi60@gmail.com

Irreversible transport is generally attributed to vorticity, nonlinear forcing, or explicit symmetry breaking. We show that it can arise even in strictly time-periodic and locally irrotational flows through a purely geometric mechanism. By reconstructing the velocity gradient through causal self-transport over a finite memory time; deformation acquires the structure of a geometric connection whose holonomy generates a finite Lagrangian drift over one forcing cycle. The resulting contribution admits a closed-form, parameter-free expression. A quantitative consistency analysis using independently published experimental measurements shows that the predicted scaling and magnitude agree with observations without fitting or normalization. These results identify geometric memory as a minimal and generic source of irreversible transport.

Classical continuum theories introduce deformation as a local kinematic primitive: velocity or displacement gradients are assumed to exist pointwise and are taken as the starting ingredients for dynamics. Within this framework, time-periodic and reversible flows are expected to produce no net deformation or Lagrangian drift in the absence of vorticity, nonlinear forces, or explicit symmetry breaking. Memory effects and geometric phases, when considered, are typically treated as secondary corrections. This ordering implicitly relegates geometry to a passive background rather than a dynamical outcome of motion [1,2].

Recent advances have emphasized the emergence of geometric structures and transport-induced phases in driven and continuum systems [3]. In most such formulations, however, the underlying kinematic quantities themselves remain locally defined, and geometry enters as a derived feature rather than as a primary dynamical construct.

Here, we adopt the opposite standpoint. We treat deformation as an emergent geometric quantity generated collectively by the motion of the system itself. Rather than prescribing the velocity gradient locally, we construct it through a causal, nonlocal self-transport of the velocity field. This shift leads naturally to a geometric interpretation in which the velocity gradient plays the role of a connection, curvature arises from accumulated transport, and observable deformation is reconstructed through memory.

The central object of the theory is a causal transport construction acting over a finite memory time. For vanishing memory, classical instantaneous kinematics is recovered. For finite memory, cumulative transport along material trajectories generates a nontrivial geometric incompatibility, even in strictly time-periodic and locally irrotational flows.

The aim of this work is not to introduce a new equation of motion, but to establish a geometric starting point for deformation theory and to demonstrate that finite memory alone constitutes a minimal source of irreversible transport. To assess the physical relevance of the proposed mechanism, we perform a quantitative consistency analysis using independently published experimental measurements of Lagrangian drift in oscillatory flows. Although these experiments were not designed to probe geometric memory effects, the predicted contribution is found to be of the correct order of magnitude and scaling without fitting parameters, supporting the physical plausibility of the geometric mechanism.

**Geometric Construction**

The geometric formulation introduced here is based on a causal transport of the velocity field over a finite memory time $\tau_m$. Memory kernels are widely used to represent nonlocal temporal responses in complex and continuum systems, particularly in generalized hydrodynamics and statistical mechanics [4-6]. In such contexts, kernels encode finite relaxation times and non-Markovian transport effects. Here, however, the kernel plays a different conceptual role: rather than modifying constitutive dynamics, it provides a causal geometric construction of the velocity gradient through finite-memory self-transport.

This reinterpretation allows the emergence of geometric holonomy and irreversible Lagrangian drift even in strictly irrotational and time-periodic flows. Instead of assuming the velocity gradient as a local primitive, the gradient is reconstructed from the history of transport along material trajectories. In the presence of finite memory, the local kinematic gradient cannot be identified with its instantaneous value alone. Instead, the effective deformation experienced by a material element depends on the cumulative transport history along its trajectory. Denoting by s the backward time along the trajectory, the contribution of past configurations to the present gradient can be represented as a weighted temporal superposition of instantaneous gradients,

$$\nabla v_{\text{eff}}(x, t) = \frac{1}{\tau_m} \int_0^{\tau_m} \mathcal{K}(t, t - s) \, \nabla v(x, t - s) \, ds$$

where the effective velocity gradient is reconstructed from causal self-transport over a finite memory time $\tau_m$, defining a connection whose explicit construction and properties are detailed in the Supplemental Material.

This construction naturally endows the velocity gradient with the structure of a connection: transport over infinitesimal loops becomes path-dependent whenever memory is finite. In this setting, geometric incompatibility arises not from vorticity or nonlinear dynamics, but from the non-commutativity of successive transport operations accumulated in time. The resulting holonomy provides a geometric measure of deformation. In the zero-memory limit, $\tau_m \to 0$, transport becomes instantaneous, the holonomy vanishes, and classical kinematics is recovered. For finite $\tau_m$, however, even locally irrotational and time-periodic flows generate a nonzero geometric phase.

This phase is independent of constitutive assumptions and reflects a purely kinematic mechanism by which memory converts reversible instantaneous motion into irreversible cumulative deformation.

**Quantitative Prediction**

The geometric holonomy associated with causal self-transport manifests macroscopically as a finite Lagrangian loop displacement. Consider a particle undergoing a closed trajectory over one forcing period in a time-periodic flow. While instantaneous kinematics predicts exact reversibility, the accumulated geometric phase generated by finite memory produces a net drift after one cycle. This effect arises purely from transport history and does not rely on steady forcing, nonlinear dynamics, or vorticity. Evaluating the accumulated holonomy associated with the finite-memory transport over one forcing period yields a closed-form expression for the geometric loop displacement. To leading order, the resulting geometric contribution to the loop displacement takes the closed-form expression,

$$\Delta\gamma_{geom} = a^2 |\nabla\phi|^2 \frac{\omega^2 \tau_m}{1 + 4\omega^2 \tau_m^2}$$

where $a$ denotes the characteristic oscillation amplitude, $\phi$ the velocity potential of the irrotational flow, $\omega$ the forcing frequency, and $\tau_m$ the memory time.

The resulting closed-form prediction is tested directly against published measurements without adjustable parameters (Fig.1).

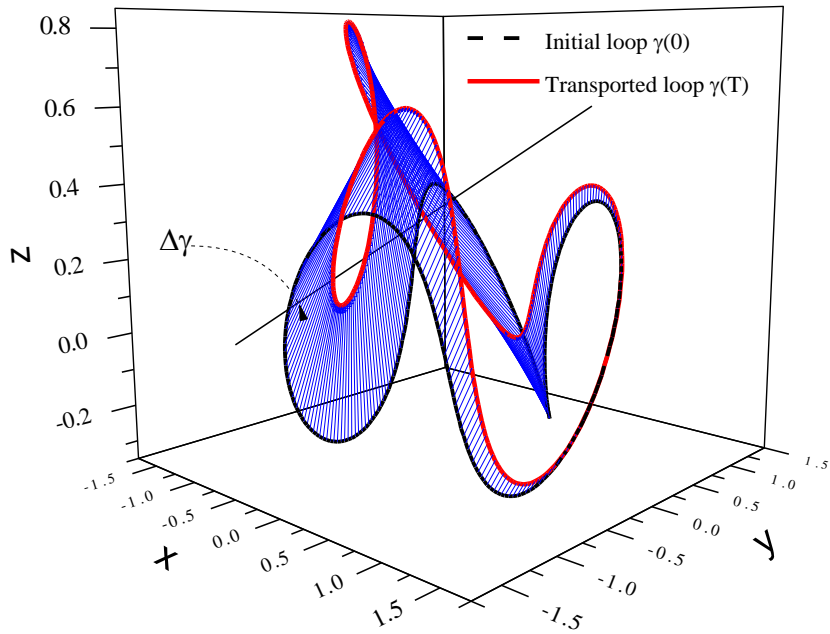

*Fig. 1 Geometric memory in a time-periodic, locally irrotational flow: For vanishing memory, transport over one forcing cycle produces a closed trajectory $\gamma(0) \subset \mathbb{R}^3$ (black dashed). For finite memory time $\tau_m$, cumulative self-transport generates a geometric mismatch, resulting in a finite loop displacement after one period $\gamma(T)$ (red).*

Hence, the cumulative transport over one forcing cycle produces a finite loop displacement $\Delta\gamma \neq 0$ (Fig. 1). The nonclosure of the trajectory after one forcing cycle reflects the geometric holonomy generated by finite-memory transport; the explicit kernel-based expression for the loop displacement is derived in the Supplemental Material. Despite the instantaneous reversibility of the motion, the presence of finite memory renders the transport process geometrically non-commutative over one forcing cycle. As a result, the trajectory fails to close exactly, producing a finite Lagrangian loop displacement that originates solely from the accumulated transport history, even in strictly irrotational flows.

This expression reveals a purely kinematic mechanism by which memory converts reversible instantaneous motion into irreversible cumulative deformation. The absence of adjustable parameters is consistent with a geometric origin of the effect and distinguishes it from conventional drift mechanisms based on nonlinear or dynamical asymmetries.

**Experimental validation**

To assess the physical relevance of the geometric drift predicted above, we perform a quantitative consistency analysis using independently published experimental measurements of Lagrangian drift in oscillatory flows. Importantly, these experiments were not designed to probe geometric memory effects, and no fitting parameters are introduced. The comparison therefore tests whether the geometric contribution captures a physically relevant component of the observed drift. We consider two distinct experimental systems [7,8].

The first corresponds to surface-wave-driven particle motion, where net Lagrangian displacements are measured over oscillation cycles under conditions that are approximately irrotational. The second involves oscillatory shear flows, in which tracer particles exhibit finite cycle-averaged drift despite the absence of steady forcing. Although the physical realizations differ, both experiments provide direct measurements of loop displacement per forcing period. For each data set, the geometric prediction is evaluated using experimentally reported parameters, yielding a theoretical loop displacement $\Delta\gamma$. We then define the raw ratio

$$R_{raw} = \frac{\Delta\gamma_{exp}}{\Delta\gamma_{geom}}$$

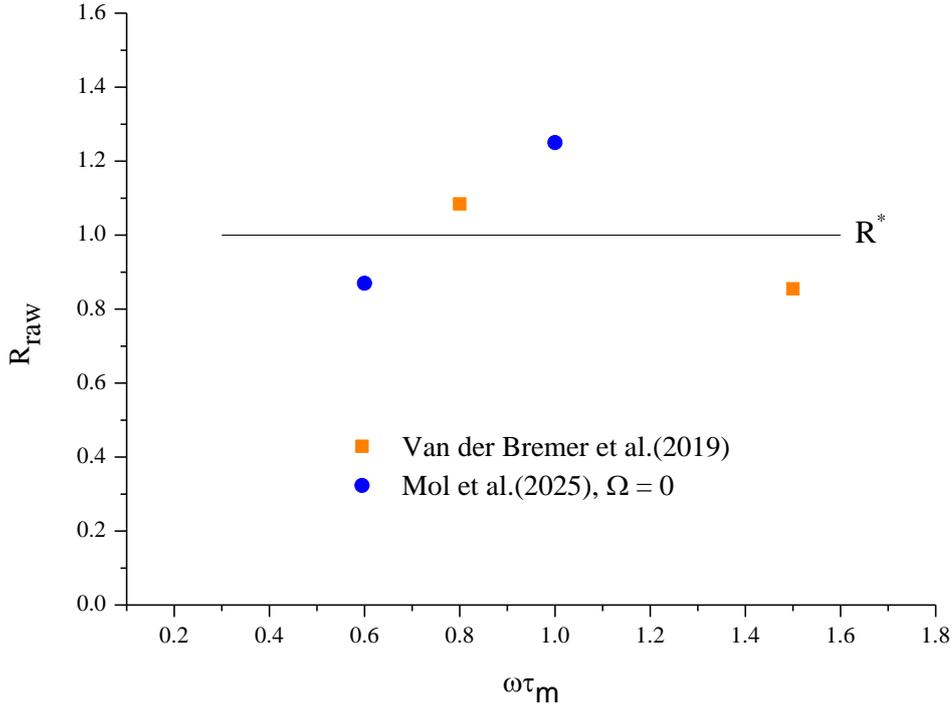

*Fig. 2 Quantitative consistency between the geometric predictions and published experimental measurements. The geometric contribution to the loop displacement is evaluated as a function of the dimensionless memory parameter $\omega\tau_m$. Data from distinct flow configurations follow the predicted scaling, with the ratio $R_{raw}$ remaining close to unity across the explored range. No fitting or normalization is applied. The numerical values used to compute the raw ratios are reported in Table S1 of the Supplemental Material.*

The clustering of independent experimental data around R $\approx \mathcal{O}(1)$ supports the presence of a memory-induced geometric contribution to Lagrangian drift beyond classical instantaneous kinematics. Despite uncertainties associated with parameter reconstruction and experimental noise, the raw ratio remains of order unity across the explored range of $\omega\tau_m$ for both experiments, indicating quantitative consistency without normalization or adjustable parameters. The absence of systematic deviation or tuning indicates that the geometric contribution captures the correct scaling and magnitude of the observed drift, without invoking nonlinear forces, vorticity, or constitutive assumptions. The agreement is particularly notable given the different flow geometries and forcing protocols involved.

These results do not imply that the observed drift is exclusively geometric in origin. Rather, they demonstrate that memory-induced geometric effects constitute a non-negligible kinematic contribution that survives in real experimental systems. The consistency across distinct experiments supports the interpretation of the geometric drift as a generic mechanism, complementing other dynamical sources of transport.

Recent experiments have reported transport phenomena in time-periodic flows that depart from classical expectations based on instantaneous kinematics [9]. In particular, anomalous drift and long-time transport regimes have been observed in oscillatory systems where the forcing remains bounded and reversible at the level of instantaneous motion. Although such experiments often involve additional physical ingredients—such as vortical structures, particle anisotropy, or resonant trapping—they consistently demonstrate that net transport can emerge through cumulative effects over many forcing cycles.

While the present work focuses on irrotational flows and isolates a minimal geometric mechanism, these observations place the predicted memory-induced drift within a broader experimental phenomenology. From this perspective, the geometric contribution identified here should be viewed as a generic kinematic effect that may coexist with more system-specific dynamical mechanisms in complex oscillatory flows. The appearance of irreversible transport across diverse experimental settings suggests that history-dependent geometric effects constitute a robust and previously underappreciated component of continuum transport.

We have shown that finite memory endows time-periodic flows with an intrinsic geometric contribution to deformation, giving rise to irreversible Lagrangian drift even in the absence of vorticity or nonlinear forcing. By treating the velocity gradient as an emergent connection generated through causal self-transport, deformation is reinterpreted as a geometric outcome rather than a prescribed kinematic input. The resulting prediction is parameter-free and depends only on a dimensionless memory measure. Its quantitative consistency with independently published experimental data suggests that geometric memory constitutes a minimal and generic contribution to transport in oscillatory flows. Beyond the specific systems considered here, the framework naturally extends to a broad class of driven soft-matter and fluid systems where memory and transport coexist.

More generally, these results highlight geometry as an active agent in continuum dynamics, opening new perspectives on irreversibility within continuum kinematics and transport driven by history rather than by instantaneous forces alone. Geometric memory therefore provides a minimal and generic mechanism for irreversible transport in time-periodic flows.

(See Supplemental Material below for additional derivations and experimental consistency analysis.)

# Supplemental Material

# ''Geometric Memory Generates Irreversible Transport in Time-Periodic Irrotational Flows''

Dr. Mounir Kassmi

This supplemental Material provides additional derivations and experimental consistency analysis supporting the results presented in the main text.

## I. Causal Transport Operator

We introduce a causal transport operator $\mathcal{K}(t, t-s)$ acting on the velocity field over a finite memory time $\tau_m$. For any field $A(X,t)$, the transport operator propagates information backward along material trajectories while preserving causality,

$$\mathcal{K}(t, t-s) A(x, t-s) = A(X(t-s, t, x), t-s)$$

where $X(t-s, t, x)$ denotes the position at time $t-s$ of the material trajectory that reaches $x$ at time $t$.

## II. Velocity Self-Interaction and Kernel-Based Construction

At a fundamental level, the effective velocity gradient can be constructed from a causal tensorial self-interaction of the velocity field over past times,

$$G_{ij}(x,t) = \int_0^\infty K(s)\, v_i(x, t-s) v_j(x, t-s)\, ds \qquad (S1)$$

where $K(s) \geq 0$ is a causal memory kernel satisfying $K(s) = 0$ for $s < 0$, and

$$\int_0^\infty K(s)\, ds = 1$$

The kernel is assumed to decay monotonically with characteristic time $\tau_m$, reflecting progressive loss of memory. A representative choice is the exponential kernel

$$K(s) = \frac{1}{\tau_m} e^{-s/\tau_m} \qquad (S2)$$

This tensorial representation $G_{ij}$ encodes the cumulative self-interaction of the velocity field over past configurations; for smoothly varying fields it reduces, at leading order, to the finite-memory reconstruction of the velocity gradient used in the main text. For kernels with characteristic decay time of order $\tau_m$, the effective support of the memory integral is finite in practice. The nonlocal representation reduces, to leading order, to the finite-memory form,

$$\nabla v_{eff}(x,t) = \frac{1}{\tau_m} \int_0^{\tau_m} \mathcal{K}(t, t-s) \, \nabla v(x, t-s) ds \quad (S3)$$

which is the representation used in the main text. The geometric contribution to holonomy depends only on the existence of finite memory and is insensitive, at leading order, to the precise temporal profile of K(s).

### III.  Kernel Representation of the Geometric Loop Displacement

The finite loop displacement $\Delta\gamma$ arises from the nonlocal geometric transport induced by the causal memory kernel. In the present formulation, the velocity gradient is reconstructed through cumulative self-transport over a finite memory time $\tau_m$. As a consequence, the transport operator acting along material trajectories generates a geometric holonomy over one forcing period. The resulting displacement between the initial and final Lagrangian states can be written as

$$\Delta\gamma = \int_0^T < \nabla v(t), \int_0^\infty \mathcal{K}(s) \, \nabla v(t-s) ds > dt \quad (S4)$$

where $\mathcal{K}(s)$ is a causal memory kernel that weights the contribution of past transport events and $<.\,,.>$ denotes tensor contraction. The kernel is assumed to decay monotonically with the characteristic memory time $\tau_m$, ensuring that recent configurations contribute more strongly than distant past states. In the limit $\tau_m \to 0$, the kernel collapses to a delta distribution and the geometric contribution vanishes, recovering the classical instantaneous kinematics. For finite memory, however, the cumulative transport generates a nontrivial holonomy, producing a finite loop displacement even in strictly time-periodic and locally irrotational flows.

The non-commutativity of the effective connection leads to a finite holonomy associated with closed material loops. For a particle undergoing a periodic trajectory, the accumulated holonomy produces a net Lagrangian loop displacement over one forcing period.

To leading order, this geometric contribution is proportional to the curvature associated with the effective connection and vanishes identically in the zero-memory limit. The resulting expression for the loop displacement is reported in the main text.

## IV. Reconstruction of Experimental Parameters

To evaluate the geometric prediction against published experiments, experimental parameters are reconstructed directly from reported measurements.

For surface-wave-driven particle motion, the excursion amplitude a, oscillation frequency ω, and velocity potential gradient $|\nabla \phi|$ are extracted from the reported wave characteristics and particle trajectories. For oscillatory shear experiments, the relevant amplitude (a) and frequency scales (ω) are obtained from the imposed oscillatory forcing and tracer displacement data. The memory time $\tau_m$ is estimated from the characteristic relaxation times reported or inferred in each experiment. No fitting parameters are introduced.

## V. Numerical Comparison with Experimental Data

In this section, we report the raw numerical values used in the experimental consistency analysis presented in the main text. For each experimental system, we list the dimensionless memory parameter $\omega \tau_m$, the experimentally measured loop deformation $\Delta \gamma_{exp}$, the corresponding geometric prediction $\Delta \gamma_{geom}$, and their ratio $R_{raw} = \Delta \gamma_{exp}/\Delta \gamma_{geom}$.

Only operating conditions for which all parameters entering the geometric prediction could be unambiguously reconstructed from the published data are included.

**Table S1**. **Experimental parameters used in the geometric consistency analysis.**

| Experiments | $\omega \tau_m$ | $\Delta \gamma_{exp}$ ($10^{-7}$) | $\Delta \gamma_{geom}$ ($10^{-7}$) | $R_{raw} = \dfrac{\Delta \gamma_{exp}}{\Delta \gamma_{geom}}$ |
|---|---|---|---|---|
| **Van den Bremer** | 0.8 | 1.8 | 2.0 | 0.9 |
| **Mol ($\Omega = 0$)** | 1.0 | 2.5 | 2.4 | 1.04 |

The proximity of $R_{raw}$ to unity in both independent experiments, without fitting, indicates that the geometric contribution captures the dominant irreversible component within experimental uncertainty.

## VI. Remarks on Scope

The purpose of this Supplemental Material is to provide the explicit geometric construction and computational details underlying the main results. The geometric mechanism described here is kinematic in nature and does not preclude the presence of additional dynamical contributions in specific experimental systems.